\documentstyle[epsf]{article}
\newcommand{\be}{\begin{equation}}
\newcommand{\ee}{\end{equation}}

\begin{document}
\title{
Gravitating monopoles in SU(3) gauge theory}
\author{{\large Y. Brihaye}$^{\ddagger}$,
and {\large B.M.A.G Piette}$^{\diamond}$\\ \\
$^{\ddagger}${\small Physique-Math\'ematique, Universite de 
Mons-Hainaut, Mons, Belgium}\\ \\
$^{\diamond}${\small  Department of Mathematical Sciences, University of 
Durham, Durham DH1 3LE United-Kingdom}}

\date{\today}

\maketitle
\begin{abstract}
We consider the Einstein-Yang-Mills-Higgs equations for an SU(3)
gauge group in a spherically symmetric ansatz. Several properties
of the gravitating monopole solutions are obtained an compared with
their SU(2) counterpart.  
\end{abstract}

\section{Introduction}
During the last few years a lot of attention was devoted to
classical solutions of conventional field theories coupled to gravity,
this after the remarkable work of Bartnik-McKinnon \cite{bm}.
One of the most fashionable field theories in which static, regular,
finite energy solution exist is the Georgi-Glashow model:
an SU(2)-Yang-Mills gauge theory coupled to a triplet of scalar 
fields and with a Higgs potential breaking the symmetry.
This model admits  topological solitons among its classical solutions~:
the celebrated SU(2)-t'Hooft-Polyakov monopole \cite{thooft,polyakov}
and the multi-monopoles.
Soon after the construction of the first monopole solution in SU(2),
the classification of its counterparts in larger groups has been
investigated, e.g. \cite{cfno,bn,tom,km};  
with an Higgs-field multiplet in the adjoint
representation of the gauge group.

Several years ago, it was shown that SU(2)-gravitating magnetic monopoles, 
as well as non-abelian black holes, exist 
in the Georgi-Glashow model coupled to gravity 
\cite{lnw,bfm1,bfm2}. For instance,
for a fixed value of the Higgs boson mass,
the gravitating monopoles 
exist up to a critical value $\alpha_{\rm c}$ of the parameter 
$\alpha$ (the ratio of the vector meson mass 
to the Planck mass). 
At the critical value $\alpha_{\rm c}$, a critical solution with a 
degenerate horizon is reached.
In particular, for small values of the Higgs boson mass,
the critical solution where an horizon first appears
corresponds to an extremal Reissner-Nordstr\o m (RN) solution
outside the horizon while it is non-singular inside.

Recently, Lue and Weinberg  reconsidered the SU(2) equations of the
self-gravitating magnetic monopoles
and discovered an insofar unsuspected phenomenon \cite{lw}.
Indeed,  for large enough values of the Higgs boson mass,
the critical solution is an extremal black hole with non-abelian hair
and a mass less than the extremal RN value.
This new bifurcation pattern was checked to be also present for black-holes
solutions \cite{bhk} and it is  natural to ask if it is
also present for gauge groups larger than SU(2).

In this  paper, we consider the SU(3)-generalisation of the Georgi-Glashow
model that we couple to gravity. We establish the equations of motion
of this theory in the spherically symmetric ansatz of the fields
and we construct numerically the corresponding magnetic monopole solutions.
We find that the main features of the SU(2)-gravitating monopoles
\cite{bfm1,lw} also hold for the gauge group SU(3). 
The paper is organised as follows~: in Sect. 2 we describe the Lagrangian,
the spherically symmetric ansatz and the corresponding field equations.
The main properties of the SU(2)-gravitating monopole are summarised in
Sect. 3. Our results concerning the solutions with SU(3) as gauge group
are discussed in Sect. 4 and illustrated by three figures.

\section{The equations}

 The SU(3) Einstein-Yang-Mills-Higgs action can be constructed 
in analogy with the  corresponding SU(2) one 
\cite{lnw,bfm1,bfm2}; it is described by the action
\be
      S = \int \ d^4x \ \sqrt{-g} ( {\cal L}_E + {\cal L}_{YMH})
\label{action}
\ee
with ${\cal L}_E = 16 \pi G \cal R$ (${\cal R}$ is the scalar curvature,
$G$ is  Newton's constant) and, for the matter fields
\be
    {\cal L}_{YMH} = - \frac{1}{2}{\rm Tr} F_{\mu \nu} F^{\mu \nu}
                            + \frac{1}{2}{\rm Tr} D_{\mu}\phi D^{\mu} \phi
                            - V(\phi) \ \ .
 \ee
 The usual definitions of the covariant derivative and field strengths are
 used in the above formula~:
\be
     F_{\mu \nu} = \partial_{\mu} A_{\nu}- \partial_{\nu}A_{\mu}
                   - i \ e \ [A_{\mu},A_{\nu}]
\ee
\be
         D_{\mu}\phi = \partial_{\mu}\phi - i \ e \ [A_{\mu},\phi]
\ee
with
\be
       A_{\mu}= \frac{1}{2} A_{\mu}^a \lambda_a \ \ \ \ , 
               \ \ \ \phi = \phi^a \lambda_a 
\ee
($\lambda_a, a=1,\dots,8$, 
are the Gell-Mann matrices normalised according to 
${\rm tr}(\lambda_a \lambda_b) = 2 \delta_{a,b}$).
The Higgs potential $V(\phi)$
is the most general renormalisable potential 
which breaks the SU(3) symmetry
\cite{bur}
\be
      V(\phi) = p(\upsilon^2 - \frac{1}{2}{\rm Tr} \phi^2)^2 
               +q( \upsilon^2 \frac{1}{2}{\rm Tr} \phi^2 
                  + \upsilon \sqrt{3}{\rm Tr} \phi^3  
                  - \frac{1}{2}({\rm Tr} \phi^2)^2 
                  + \frac{3}{2}{\rm Tr} \phi^4) 
\ee
$p,q$ are positive constants. 
For generic values of the constants $p,q$, the breakdown of the
SU(3)-symmetry is such that the expectation value $\langle \phi \rangle$
possesses two equal eigenvalues (i.e. the eigenvalues
of this matrix are proportional to those of $\lambda_8$).

In this pattern, we are left with four massive vector bosons
corresponding to $A^a_{\mu}$, $a=4,5,6,7$ with equal masses $M_W$
 and four scalar fields.
 The masses of these fields are given by
\be
\label{mass}
       M_W^2 = \frac{3}{4} e^2 \upsilon^2 \ \ \ , \ \ \ 
       M_1^2 = M_2^2 = M_3^2 = 18 q \upsilon^2  \ \ \ , \ \ \
       M_8^2 = 2(4p+q) \upsilon^2
\ee
In contrast, in an SU(2) gauge theory with potential 
$V(\phi) = p(\upsilon^2 - (1/2) {\rm Tr} \phi^2)^2$, there are two
vector bosons of equal masses $M_W$ and one scalar of mass $M_H$, with
\be
       M_W^2 = e^2 \upsilon^2 \ \ \ , \ \ \ M_H^2 = 8p \upsilon^2
\ee

For later use we define the following combinations of the coupling constants
\begin{eqnarray}
       &b \equiv (\frac{e M_H}{2 M_W})^2 = 2p    &{\rm for \ SU(2)} \\
       &b \equiv ( \frac{e M_8}{2 M_W})^2 = \frac{2(4p+q)}{3} \ \ , \ \ 
        R \equiv (\frac{M_3}{M_8})^2 = \frac{9q}{4p+q}   &{\rm for \ SU(3)}
\end{eqnarray}
(remark that $0 \leq R \leq 9$).
In relation with Newton's constant $G$, we define 
\be
\label{alphaa}
       \alpha^2 = 4 \pi G \upsilon^2 \ \ \ , \ \ \ \ 
       a = 8 \pi \frac{M_W^2}{e^2 M_{pl}^2}  \ \ ,
\ee
so that $a=2 \alpha^2$ and $a=(3/2) \alpha^2$ respectively for SU(2)
and SU(3).

 In order
to obtain static, spherically symmetric and globally regular solutions,
we use Schwarzschild like coordinates for the metric
\begin{equation}
\label{metric}
ds^2=
  -A^2N dt^2 + N^{-1} dr^2 + r^2 (d\theta^2 + \sin^2\theta \ d\phi^2)
\ . \end{equation}
The crucial step to obtain the flat monopole is the embedding of the su(2)
algebra (whose generators will be labelled by $T_1,T_2,T_3$)
into the su(3) algebra \cite{cfno}.
Here we will use the embedding
\be
(T_1, T_2, T_3) = (\frac{1}{2} \lambda_1,\frac{1}{2} \lambda_2, 
                  \frac{1}{2} \lambda_3)
                \equiv \frac{1}{2} \vec \tau
\ee
for which the monopole with the lowest classical energy can be constructed.  
Then, we use the spherically symmetric ansatz for the spatial 
components of the gauge field and for the
Higgs fields (see e.g. \cite{km})
\begin{equation}
\vec A_\theta =  -\vec e_\phi \frac{1- K(r)}{e  r} \ , \ \ \
\vec A_\phi =   \vec e_\theta \frac{1- K(r)}{e r} \sin \theta
\ , \ \ \   \phi =  
\upsilon(H_1(r) \vec \tau \vec e_r + H_2(r) \lambda_8) \ , 
\end{equation}
with the standard unit vectors $\vec e_r$, $\vec e_\theta$ and $\vec e_\phi$.
It is  convenient to use the dimensionless variable 
$x=e \upsilon r$ and to define  the mass function $\mu(x)$ by means of
$N(x) = 1-2 \mu(x)/x$ 

With these ansatz and definitions, the classical equations of 
the Lagrangian (\ref{action})
reduce to a system of five differential equations.
For the functions parametrising the metric we have 
\begin{eqnarray}
\mu'&=&\alpha^2 \Biggl(  N K'^2 + \frac{1}{2} N x^2 (H_1'^2 + H_2'^2)
\nonumber\\
 & &\phantom{ \alpha^2 \Biggl( }
   + \frac{(K^2-1)^2}{2 x^2} + H_1^2 K^2
   + p x^2 (H_1^2 + H_2^2-1)^2  + q x^2(H_1^2(1+2H_2)^2 
+ (H_2 + H_1^2 - H_2^2)^2) \Biggr)
\ , \label{eqmu} \end{eqnarray}

and
\begin{eqnarray}
 A'&=&\alpha^2 x \Biggl(
    \frac{2 K'^2}{x^2} + H_1'^2 + H_2'^2 \Biggr) A
\ , \label{eqa} \end{eqnarray}
(the prime indicates the derivative with respect to $x$). 
For the functions related to the matter fields, we obtain
\begin{eqnarray}
(A N K')' = A K \left( \frac{K^2-1}{x^2} + H_1^2 
 \right)
\ , \label{eqk} \end{eqnarray}
\begin{equation}
(x^2 A N H_1')' = A \left( 2 H_1 K^2 + x^2 \frac{\partial V}{\partial H_1}
\right) 
\ , \label{eqh1} \end{equation}
and
\begin{eqnarray}
( x^2 A N H_2')' = A \left(  x^2 \frac{\partial V}{\partial H_2}
    \right)
\ . \label{eqh2} \end{eqnarray}

The solutions of these equations can be studied for different 
values of the
physical parameters $\alpha^2 , p, q$.
They are characterised, namely, 
by the mass $\mu_{\infty}$ (the asymptotic value of the function 
$\mu(x)$) and by their inertial mass $M = \mu_{\infty}/ \alpha^2$, 
which, after multiplication by
$4 \pi \upsilon / e$ gives the ADM mass.

The regularity of the solution at the origin (including the function
$N(x)$), 
the integrability of the mass function $\mu(x)$ 
and the requirement that the metric (\ref{metric})
asymptotically approaches the Minkowski metric 
lead to the following set of boundary conditions
\be 
\mu(0) = 0 \ \ , \ \ K(0) = 1 \ \ , \ \ H_1(0) = 0 \ \ , \ \ H'_2(0) = 0
\ee
\be
A(\infty) = 1 \ \ , \ \ K(\infty)=0 \ \ , \ \ H_1(\infty) =  
         \frac{\sqrt{3}}{2} \ \ ,
\ \ H_2(\infty) =- \frac{1}{2}  \ \ 
\ee
which complete the mathematical problem posed by 
Eqs. (\ref{eqmu})-(\ref{eqh2}).
In fact, alternative boundary conditions for the matter functions
$K,H_1,H_2$ are also possible \cite{km}
but we will restrict ourselves
to the ones above which, in the flat limit, corresponds to the monopole
with the lowest energy.

In the absence of gravity (i.e. $\alpha=0$, $N=A=1$),
the first two equations are trivial and  
the SU(3)-monopole solutions are recovered.
They were studied numerically in some details in
\cite{km}. 

Setting $H_2=0$, $q=0$ in Eqs. (\ref{eqmu})-(\ref{eqh1}) 
(Eq.(\ref{eqh2}) is then trivial) leads to the equations of the
SU(2) gravitating monopole.
These were studied in details in \cite{bfm1,bfm2}. 
In particular,
it was shown that the non-Abelian gravitating dyon
bifurcates at some critical value
$\alpha =\alpha_{c}$ into an extremal Schwarzschild
solution with
\be
      N(x) = \frac{(x - \alpha)^2}{x^2} \ \ , \ \ A(x) = 1
\label{rn1}
\ee      
\be
      K(x) = 0 \ \ , \ \ H(x) = 1 \ \  
\label{rn2}
\ee
defined on $x \in [\alpha, \infty]$.
The dependence of the critical value $a_{c}$ 
(with $a$ defined in Eq. (\ref{alphaa}))
on the parameter $b$ is presented in Fig. 1 by the line
with the circles.

\begin{figure}[htbp]
\unitlength1cm \hfil
\begin{picture}(8,8)
 \epsfxsize=8cm \epsffile{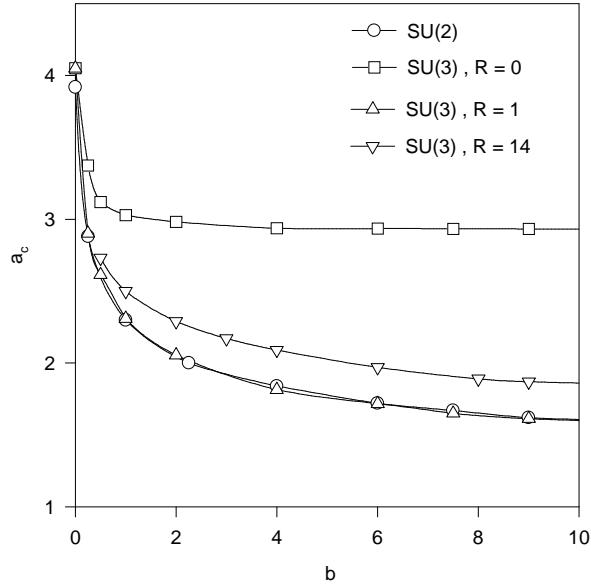}
\end{picture}
\caption{The critical value $a_c$ of the SU(2)-monopoles 
(bullets line) and of the SU(3)-monopoles (squares and triangles lines)
for the small values of the parameter $b$.} 
\end{figure}

\section{SU(2)-Gravitating monopole}

Since the purpose of this paper is to compare the properties 
of the gravitating monopoles available in SU(2) and in SU(3),
we first briefly recall how the magnetic monopole solutions
approach the critical solution
when the vector boson mass, i.e. $a$, is varied
while the ratio of the Higgs boson mass
to the vector boson mass, i.e.~$b$, is kept fixed

Recently, Lue and Weinberg \cite{lw} realized that there are two
regimes of $b$, each with its own type of critical solution.

In the first regime, $b$ is small ($b < 25$)
and the metric function $N(r)$ 
of the monopole solutions possesses a single minimum.
As the critical solution is approached, i.e.~as $a \rightarrow a_{\rm cr}$,
the minimum of the function $N(r)$
decreases until it reaches zero at $r = r_0$.
The limiting solution
corresponds to an extremal RN black hole solution with horizon radius
$r_h = r_0$ and unit magnetic charge
for $r \ge r_0$.
Consequently, also the mass of the limiting solution
coincides with the mass of this extremal RN black hole.
However, the limiting solution is not singular
in the interior region $r < r_0$.
We will call this type of limiting approach the RN-type.

In the second regime, $b$ is large
and the metric function $N(r)$ 
of the monopole solutions develops a second minimum
as the critical solution is approached.
This second minimum arises for a value $r=r_*$
which is smaller than the first one at $r=r_0$.
Keeping $b$ fixed and increasing $a$ one observes that
the internal minimum  $N(r_*)$
decreases faster  than the external one $N(r_0)$.
Therefore, the critical solution is reached for $a=a_{cr}$
when the inner minimum reaches zero.
This solution thus possesses an extremal horizon
at $r_* < r_0$,
and corresponds to an extremal black hole with non-abelian hair
and a mass less than the extremal RN value.
Consequently, we call  this second type of
limiting approach the NA-type (non-abelian-type).

To summarise~: the non-Abelian gravitating monopoles 
exist on a portion
of the $(a,b)$ plane limited by a curve $a_{cr}(b)$. 
At a particular point $(a_{tr},b_{tr})$ on this curve,  
the separation between  the RN-type and NA-type of ending occurs.
The value $a_{tr}=3/2$ is determined algebraically in \cite{lw}, 
a numerical analysis \cite{bhk} indicates  $b_{tr} \sim 26.7$. 

\section{SU(3)-Gravitating monopole}
The numerical solutions of eqs. (\ref{eqmu} -\ref{eqh2})
strongly indicates that the bifurcation pattern of the gravitating
monopoles in SU(3) is very similar to the one in SU(2). The analysis
is rendered more involved by the presence of an additional equation
and of a second parameter in the Higgs potential.

We first discuss the case when $b$ is small.
Fixing values for $b, R$ and increasing $\alpha$, the function 
$N(x)$ develops a minimum, say $N_m$ at $x=x_m$, which becomes
deeper and deeper. 
At some critical value $\alpha_c$ we have
\be
    N_m = 0 \ \ \  , \ \ \ \alpha_c = \mu_{\infty} = x_m
\ee
characterising a bifurcation into an extremal RN black hole
with an horizon $x_h$ coinciding with $x_m$.
The limiting solution is given again by (\ref{rn1}), $K(x)=0$,
$H_1(x)= \sqrt{3}/2$, $H_2(x)= -1/2$.

This approach of the RN black hole by the gravitating monopole
is illustrated 
in Fig. 2 for $b=R=1$; in this case we find $\alpha_c \approx 1.2483$.
The figure further illustrates the evolutions of the 
mass $\mu_{\infty}$  of the solution and of the inertial mass
$M = \mu(\infty)/\alpha^2$. 
Pushing the numerical analysis up to $N_m \approx 0.0001$,
we have not gotten any evidence of a second branch of solutions.
Backbending branches, which occur in the SU(2) case
(see \cite{bfm1} pp.365, Fig. 3) likely exist too in SU(3) for lower 
values of the parameters $R$ and  $b$; we have not attempted to 
construct them.

\begin{figure}[htbp]
\unitlength1cm \hfil
\begin{picture}(8,8)
 \epsfxsize=8cm \epsffile{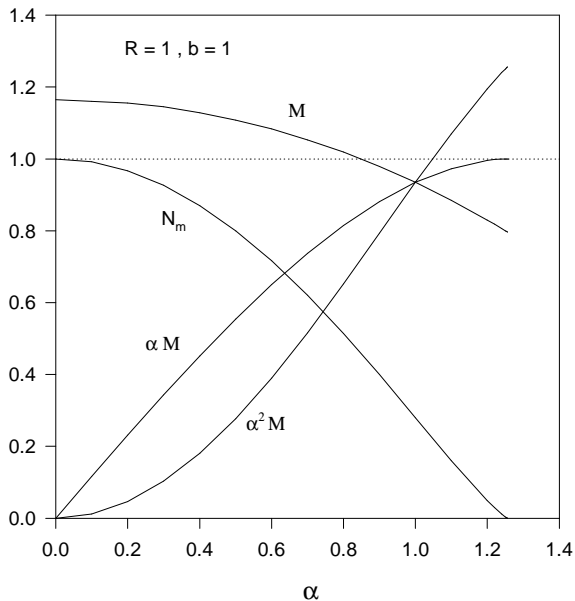}
\end{picture}
\caption{The $\alpha$ dependance of the mass $M$ and of the minimal value 
$N_m$ of $N(x)$; $N_m$ = 0 at $\alpha \approx 1.2483$.}  
\end{figure}

\begin{figure}[htbp]
\unitlength1cm \hfil
\begin{picture}(8,8)
 \epsfxsize=8cm \epsffile{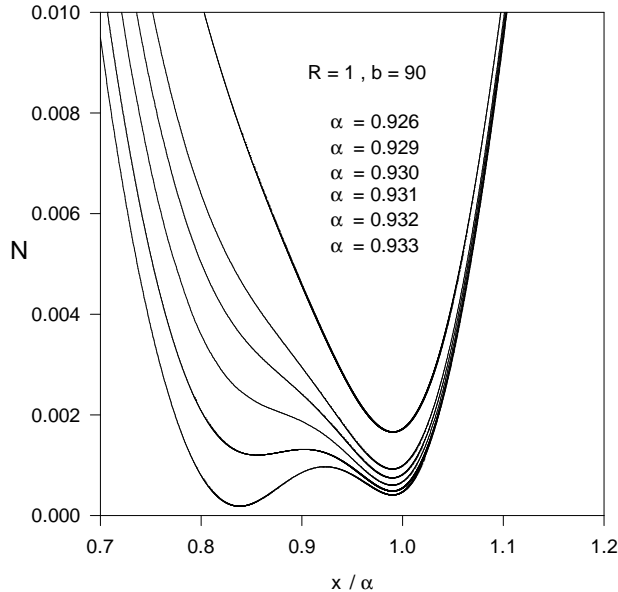}
\end{picture}
\caption{The evolution of the function $N(x/\alpha)$ at the approach 
of the critical value $\alpha_c \approx 0.9332$ for $b=90, R=1$.} 
\end{figure}

The critical value $\alpha_c$ (and correspondingly $a_c$)
off course depends on $b$ and $R$. The dependence 
of $a_c$ on the other parameters is illustrated in Fig. 1
for $R=0$ and $R=1$.
For all values covered by Fig. 1 the gravitating monopole bifurcates
into a extremal RN black hole.
As far of the horizon $x_h=x_m$ of the limiting monopole solution
is concerned, we see that , for fixed $b$, the SU(3) solution
having the largest horizon and the highest inertial mass
are those corresponding to the masslesness of the three supplementary
 Higgs particles ($M_1=M_2=M_3$ in  (\ref{mass})), i.e. to $q = 0$.


Again keeping $b$ fixed but decreasing $R$ we observed that the value 
$a_c(R)$ decreases from the value $a_c(0)$ (symbolised by
the curve with the squares in Fig. 1) and quickly reaches a minimum
for $R \approx 4.5$ (virtually overlapping the curve for $R=1$ in Fig. 1).
For fixed $b$, the value $a_c(R)$ varies only a little for $1\leq R \leq 9$;
however for $R > 9$ we observed that $a_c(R)$ raises up sensibly, as 
illustrated in Fig. 1 for $R=14$. Due to the numerical difficulties we have 
not attempted to obtain the limiting curve $a_c(R=\infty)$.


Next we considered the case when $b$ is large 
and attempted to reproduce 
the phenomenon observed by Lue-Weinberg \cite{lw} (and described
in the previous section)  for SU(3)-gravitating monopoles.
Choosing the values  $R=1$, $b=90$ we found that an inner
minimum appears for $\alpha \approx 0.932$ (while the outer
minimum has $N_m \approx 0.0007$).
Rapidly the inner minimum becomes lower than the outer one and
it is attained for $x\approx 0.78$ while $\alpha$ approaches
the critical value $\alpha_c \approx 0.9331$. 
Several profiles  illustrating the evolution of $N(x)$ 
are presented in Fig. 3. For $R=1$ and lower values of $b$,
the pattern is, qualitatively and quantitatively, very similar
to the one obtained in SU(2) and discussed at length in \cite{lw,bhk}.
However, choosing small values of the mass ratio $R$
(typically $R\sim 0.1$), we got no evidence
of this type of phenomenon.

The numerical analysis of the equation is very lengthly and our goal
is just to demonstrate that the "NA-type" of bifurcation is also
present for groups larger than SU(2), we therefore have not attempted
to refine the evaluation of $b_{tr}$ as a function of $R$.

\section{Conclusion}
The Einstein-Georgi-Glashow model constitutes a good theoretical 
laboratory for testing the properties of gravitational solitons. 
The set of solutions is particularly rich and contains  
patterns of bifurcations of several types \cite{bfm2,lw,bhk}. 
In this paper we have shown that many properties of these solutions are present
with the larger group SU(3), in particular gravitating monopoles
can bifurcate into both Abelian and non-Abelian (hairy) black holes,
according to the values of the coupling constants. This result
suggests that these two types of critical phenomenon are not
specific to SU(2) \cite{lw, bhk} and occur in SU(N) and, in particular, in the
"grand-unifying models".


\vfill


\end{document}